\documentstyle[aps,prb,twocolumn,epsf,floats]{revtex}
\draft

\begin{document}
\wideabs{
\title{Theory of combined exciton-cyclotron resonance
       in a two-dimensional electron gas:
       The strong magnetic field regime }
\author{A. B. Dzyubenko\cite{ABD}}
\address{Department of Physics, SUNY at Buffalo, Buffalo, NY 14260, USA }
\date{Phys. Rev. B {\bf 64}, 241101(R) (2001)}
\maketitle
\begin{abstract}
I develop a theory of combined exciton-cyclotron resonance (ExCR)
in a low-density two-dimensional electron gas in high magnetic fields.
In the presence of excess electrons an incident photon creates an exciton and
simultaneously excites one electron to higher-lying Landau levels.
I derive exact ExCR selection rules that follow from the existing
dynamical symmetries, magnetic translations and rotations about
the magnetic field axis. The nature of the final states in ExCR
is elucidated. The relation between ExCR and shake-up
processes is discussed. The double-peak ExCR structure for transitions
to the first electron Landau level is predicted.
\end{abstract}
\pacs{71.35.Cc,71.35.Ji,73.21.Fg}
}


Optical manifestations of many-body effects in low-dimensional
electron-hole ($e$--$h$) systems in magnetic fields
have been the focus of many experimental and theoretical studies
during the past decade. Recently, such objects as, e.g.,
artificial atoms in quantum dots and negatively charged excitons $X^-$
in quantum wells have been under intense scrutiny. \cite{EP2DS}
The surprising apparent stability of the $X^-$
(which in the dilute limit is a weakly bound state \cite{Cox}
of two electrons and one hole)
in the presence of excess electrons in strong magnetic fields,
and the relation of this stability to a many-body collective
response have been actively discussed. \cite{EP2DS,many-body}
Another interesting manifestation of many-body effects
are shake-up processes \cite{Shake} in the photoluminescence
of a two-dimensional electron gas (2DEG): After the recombination
of the $e$--$h$ pair, one electron is excited to one of
the higher Landau levels (LL's). A closely related phenomenon, combined
exciton-cyclotron resonance (ExCR) also has been identified
\cite{ExCR} in low-density 2DEG systems: Here, an incident photon
creates an exciton and simultaneously excites one electron to higher LL's.
These phenomena and the relation between them remain only partially
understood. \cite{EP2DS,Cox,many-body,Shake,ExCR}
A theory of ExCR has been developed \cite{ExCR} for weak magnetic
fields, when the magnetic length $l_B=(\hbar c/eB)^{1/2}$
is much larger than the exciton Bohr radius
$a_B=\epsilon \hbar^2/m e^2$:  $l_B \gg a_B$.
The energy positions of the ExCR spectra were obtained \cite{ExCR}
from an expansion in the unbound ``exciton+electron'' states.
The Coulomb interactions were taken into account phenomenologically
as two-particle $e$--$h$ excitonic corrections to the transition
matrix elements.
In addition, the following assumptions \cite{ExCR} were made:
(i) a strictly-2D system,
(ii) background electrons in the lowest spin-polarized
$n=0$\,$\uparrow$ LL with spins oriented parallel to the field,
(iii) low electron density $n_e$, i.e., $n_e a_B^2 \ll 1$.

In this work, I develop the theory of ExCR for the
physically interesting regime of {\em strong} magnetic fields,
$l_B \ll a_B$. Otherwise, I adopt essentially
the same assumptions (i) --- (iii), as in Ref.~\onlinecite{ExCR}.
Note that in the high-$B$ limit, the characteristic length of the
problem is $l_B$ rather than $a_B$, and condition (iii)
can be formulated in terms of the filling factor,
as $\nu_e = 2\pi l_B^2 n_e \ll 1$. In the limit of low electron
density, ExCR can be considered \cite{ExCR}
to be a three-particle resonance involving a {\em charged system}
of two electrons and one hole, $2e$--$h$, in the final state.
Importantly, there is a coupling of the center-of-mass and
internal motions for charged $e$--$h$ complexes in magnetic
fields. \cite{Simon} In order to describe the high-field ExCR,
I obtain the complete spectra of the $2e$--$h$ eigenstates in higher LL's
with a consistent treatment of the Coulomb correlations.
I exploit a recently developed scheme \cite{Dz&S_PRL,SSC} for charged
$e$--$h$ complexes in magnetic fields in which one degree of freedom is
separated while all existing dynamical symmetries, rotations about
the ${\bf B}$ axis and magnetic translations, are preserved.
This allows one to establish exact ExCR selection rules
that are applicable in {\em arbitrary\/} magnetic fields,
to derive the ExCR oscillator strengths, and to establish
the heretofore missing relation between ExCR and shake-up
processes in the dilute limit.

The Hamiltonian describing the 2D $2e$--$h$ state
in a perpendicular magnetic field ${\bf B}=(0,0,B)$ is given by
\begin{equation}
                \label{H} 
  H = \sum_{i=1,2} \frac{\hat{\bbox{\pi}}_{ei}^2}{2m_e} +
                   \frac{\hat{\bbox{\pi}}_{h}^2 }{2m_h}
      -  \sum_{i=1,2} \frac{e^2}{\epsilon |{\bf r}_i-{\bf r}_h|}
      +               \frac{e^2}{\epsilon |{\bf r}_1-{\bf r}_2|}  \, ,
\end{equation}
where $\hat{\bbox{\pi}}_j = -i\hbar \bbox{\nabla}_j -
\frac{e_j}{c} {\bf A}({\bf r}_j)$
are kinematic momentum operators
and the symmetric gauge ${\bf A} = \frac12 {\bf B} \times {\bf r}$ is used;
a weak exchange $e$--$h$ interaction, small central-cell corrections
to the Coulomb potential, and the crystal anisotropy are not relevant
for the present study and are neglected.
The exact eigenstates of (\ref{H}) form families of degenerate states;
each family is labeled by the index $\nu$ that
plays a role of the principal quantum number and
can be discrete (bound states) or continuous (unbound states forming
a continuum). \cite{Dz&S_PRL,SSC}
There is a macroscopic number of degenerate states
in each family labeled by the discrete
oscillator quantum number $k=0, 1, \ldots$.  This quantum number
is associated with magnetic translations \cite{Simon,Dz&S_PRL,SSC}
and physically describes the center-of-rotation\cite{rmrk} of the charged
complex in $B$. Each family starts with its Parent State (PS)
$ |\Psi_{k=0 \, M_z \, S_e S_h \nu}\rangle$ that has $k=0$
and the largest (for the total charge $Q<0$) possible in the family
value of the total angular momentum projection $M_z$.
Degenerate daughter states $|\Psi_{k\,M_z-k\,S_eS_h\nu}\rangle$
can be constructed iteratively \cite{Simon,Dz&S_PRL} out of the PS.
$S_e$ denotes the total spin of two electrons,
either $S_e=0$ (singlet states) or $S_e=1$ (triplet states);
$S_h$ is the spin state of the hole.
Non-interacting electrons (or holes) in $B$ can also be classified
according to this scheme. The corresponding $e$- and $h$- single-particle
wave functions $\phi^{(e)}_{n m}({\bf r})=\phi^{(h)*}_{n m}({\bf r})$
have the  factored form and are well-known in the theory of
the Fractional Quantum Hall Effect;\cite{FQHE}
$n$ is the LL quantum number, which determines the energy
$\hbar\omega_{\rm ce(h)}(n+\case{1}{2})$ with
$\omega_{\rm ce(h)}=eB/m_{\rm e(h)}c$, and
$m=0,1, \ldots $ is the intra-LL oscillator quantum number,
the single-particle version of $k$.
The angular momentum projections are $m_{\rm ze}=-m_{\rm zh}=n-m$.
In the zero LL
\begin{equation}
   \label{phi0m}
 \phi^{(e)}_{0m}({\bf r}) = \frac{1}{\sqrt{2\pi l_B^2m!}}
\left(\frac{z^{\ast}}{\sqrt{2}l_B}\right)^m
    \exp\left(-\frac{{\bf r}^2}{4l_B^2}\right)       \, ,
\end{equation}
where $z^{\ast}=x-iy$.

\begin{figure}[t]
\epsfxsize=3.0in
\epsffile{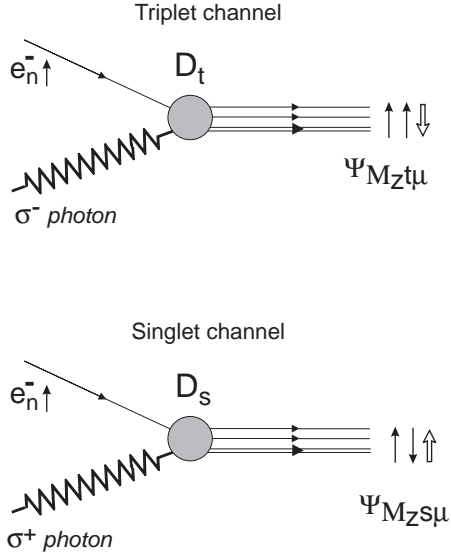}
\vspace*{3ex}
\caption{Optical transitions involving the low-density 2DEG
in the spin-polarized $n$\,$\uparrow$ LL. $\Psi_{M_z s(t) \mu}$ denotes
charged singlet $s$ (triplet $t$) $2e$-$h$ Parent States with $k=0$.
The exact selection rule is $M_z=n$.
}                \label{fig1}
\end{figure}

Let us derive the optical selection rules for ExCR.
Interband transitions with $e$--$h$ pair creation
can be described by the luminescence operator
$\hat{\cal L}_{\rm PL}= p_{\rm cv} \int \! d{\bf r} \,
\hat{\Psi}^{\dagger}_{e}({\bf r}) \hat{\Psi}^{\dagger}_{h}({\bf r})$,
where $p_{\rm cv}$ is the interband momentum matrix element.
Consider the creation of the $e$--$h$ pair in the presence
of the low-density 2DEG in the $n$-th LL (see Fig.~\ref{fig1}).
Taking into account only the three-particle correlations
in the final state, the dipole transition matrix element is
\begin{equation}
       \label{ExCR}
   D(\nu) = \langle \Psi_{k \, M_z-k \, S_e S_h \nu}| \hat{\cal L}_{\rm PL}
              |\phi^{(e)}_{nm} \rangle  \, .
\end{equation}
The oscillator quantum number is conserved\cite{Dz&S_PRL}
in dipole transitions: $m=k$. Physically this means
that the centers-of-rotation of charged complexes in the
initial and final states must coincide.
Due to the change of the Bloch wave functions, the usual selection rule
$\Delta M_z= 0$ holds for the envelope functions, thus,
$m_{\rm ze}=n-m=M_z-k$.
The combination of the two selection rules leads to
$D({\nu}) \sim \delta_{n,M_z}$,
where $M_z$ is the angular momentum projection
of the PS in the $\nu$-th family.
Therefore, in the ExCR processes involving electrons from
the $n$-th LL, the achievable final $2e$--$h$ states must have
$M_z=n$ and may belong to {\em different} final LL's.
If the 2DEG is spin-up $\uparrow$ polarized, the
photon of $\sigma^+$ ($\sigma^-$) circular polarization produces
the singlet (triplet) final states (see Fig.~\ref{fig1}).

It is illuminating to compare the ExCR processes and
photoluminescence (PL) of negatively charged excitons,
$X^- \rightarrow e^-_n + {\sl photon}$,
in which the electron is left in the $n$-th LL in the final state.
Such processes are described by the transition matrix element
$D({\nu})^* = \langle \phi^{(e)}_{nm} | \hat{\cal L}_{\rm PL}^{\dag}
| \Psi_{k \, M_z-k \, S_e S_h \nu}\rangle$
and can be considered as the inverse of ExCR.
The exact selection rule $D({\nu})^{\ast} \sim \delta_{M_z,n}$ shows that
the $X^-$ PL transition is only possible when the electron is
left in a {\em single and specific\/} LL with the number $n=M_z$.
Therefore, contrary to ExCR, no various final LL's are achievable
in the PL from any given $X^-$ state.
In this sense, the shake-up processes,\cite{Shake}
having as the final states various LL's $n=1,2,\ldots$
must be prohibited\cite{Dz&S_PRL} in the PL of the isolated
$X^-$ in a translationally invariant system in $B$.

Below I study ExCR from the zero spin-polarized
$n=0$\,$\uparrow$ LL to the first electron LL in high magnetic fields
\begin{equation}
      \label{highB}
 \hbar\omega_{\rm ce} \, , \,  \hbar\omega_{\rm ch} \gg
E_0 = \sqrt{\frac{\pi}{2}} \, \frac{e^2}{\epsilon l_B} \, .
\end{equation}
The $2e$--$h$ states can be obtained in this regime
as an expansion in LL's.
The complete orthonormal basis compatible with
both axial and translational symmetries has been constructed in
Ref.~\onlinecite{SSC} and will be denoted here as
\begin{equation}
        \label{basis}
  |n_R n_r n_h ; k m l \rangle  \, .
\end{equation}
The conserved oscillator quantum number is fixed and equals
$k$ in (\ref{basis}) and $M_z = n_R + n_r - n_h - k - m + l$;
$n_h$ is the hole LL number, $n_r$ and $n_R$ are the LL numbers of
the electron relative and center-of-mass motions,
respectively (see Ref.~\onlinecite{SSC} for details).
The permutational symmetry requires that $n_r-m$ should be even (odd) for
$S_e=0$ ($S_e=1$). Fixing $k$ in (\ref{basis}) amounts to summing
the infinite number of $e$- and $h$- states in zero LL's.
As an example, the state $| \tilde{0} \rangle= |000; 000 \rangle$ ---
the new vacuum --- has the form
\begin{eqnarray}
         \nonumber
  \langle {\bf r}_1 {\bf r}_2 {\bf r}_h | \tilde{0} \rangle &=&
    \Phi_{0}({\bf r}_1, {\bf r}_2; {\bf r}_h)
    = \frac{1}{\sqrt{2}\,(2\pi l_B^2)^{3/2}} \\
   \label{vacuum}
    & \times & \exp \left( -\frac{{\bf r}_1^2+{\bf r}_2^2+{\bf r}_h^2-
        (z^{\ast}_1 + z^{\ast}_2) z_h }{4 l_B^2} \right) \,
\end{eqnarray}
and is a coherent \cite{SSC} $e$--$h$ state [cf.\ Eq.~(\ref{phi0m})].
In what follows I consider only the PS's
$| \Psi_{M_z s(t) \nu}\rangle$, where
the quantum numbers  $k=0$ and $S_h$ are omitted for brevity
and $s$ ($t$) denotes the singlet $S_e=0$ (triplet $S_e=1$)
electron spin state.

Neglecting mixing between LL's (the high-field limit),
the triplet $2e$--$h$ states in the first electron and zero hole LL,
$| \Psi_{M_z t \nu}^{(10)}\rangle$, can be obtained as the expansion
\begin{eqnarray}
        \label{Psi10}
    | \Psi_{M_z t \nu}^{(10)} \rangle  & = &
     \sum_{l=0}^{\infty} \sum_{m=0}^{\infty}
        \alpha_{M_zt\nu}(2m,l) \, |010 ; 0 \, 2m \, l \rangle   \\
         \nonumber
    & + &
     \sum_{l=0}^{\infty} \sum_{m=0}^{\infty}
        \beta_{M_zt\nu}(2m+1,l) \, |100 ; 0 \, 2m+1 \, l \rangle \, ,
\end{eqnarray}
where expansion coefficients
$\alpha_{M_zt\nu}(2m,l) \sim \delta_{M_z,l+1-2m}$ and
$\beta_{M_zt\nu}(2m+1,l) \sim \delta_{M_z,l-2m}$.
The Coulomb matrix elements of the Hamiltonian (\ref{H})
are calculated analytically\cite{SSC} and the
eigenspectra are obtained by numerical diagonalization of
finite matrices of order $2-5\times10^2$.
Such finite-size calculations provide very high accuracy for bound $X^-$
states and are also capable of reproducing the structure of the
three-particle continuum. \cite{SSC}
The singlet states $| \Psi_{M_z s \nu}^{(10)}\rangle$ are
obtained by using a similar procedure.

\begin{figure}[t]
\epsfxsize=3.5in
\epsffile{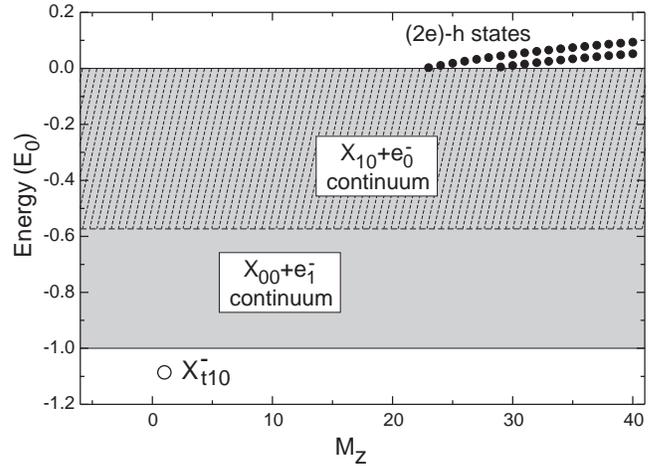}
\vspace*{2ex}
\caption{The eigenspectra of the three-particle $2e$--$h$
triplet $S_e=1$ states in the first electron LL.
The Parent States $|\Psi_{M_z t \mu}^{(10)}\rangle$
with $k=0$ are only shown. Energy is given relative to
$\hbar(\case{3}{2}\omega_{\rm ce}+\case{1}{2}\omega_{\rm ch})$
in units of $E_0=\protect\sqrt{\pi/2} \, e^2/\epsilon l_B$.}
\label{fig2}
\end{figure}

The calculated triplet $2e$--$h$ eigenspectra are presented
in Fig.~\ref{fig2}.
Filled dots above free LL's show positions of the excited
bound three-particle states, denoted as $(2e)$--$h$.
These states originate from the excited states of two electrons that
are bound in 2D because of the confining effect of the magnetic field.
Bound $(2e)$--$h$ states appear in the spectrum for relatively large
values of $M_z$, when the hole can be at a sufficiently
large distance from the two electrons.
There is also exactly one low-lying triplet bound state --- the negatively
charged magnetoexciton (MX) $X^-_{t10}$ that has
$M_z=1$ and binding energy $0.086E_0$.\cite{Dz&S_PRL,SSC}
The shaded area of width $E_0$ in Fig.~\ref{fig2} corresponds to the
three-particle continuum formed by the neutral MX $X_{00}$
($e$ and $h$ in their zero LL's) with the second electron
in a scattering state in the first LL\@.
The hatched area corresponds to the second overlapping band formed by
the states of the neutral MX $X_{10}$ ($e$ in the first and $h$
in the zero LL) with the second electron in a scattering state
in the zero LL\@. The lower continuum edge lies at the $X_{10}$
ground state energy $-0.574E_0$, which, for the isolated MX,
is achieved at a finite center-of-mass momentum
$|{\bf K}|l_B \simeq 1.19$. \cite{SSC,L&L80}
As a result, the density of $X_{10}$ states has at this energy
an inverse square-root van Hove singularity in 2D.
The continuum of the singlet $2e$--$h$ states has qualitatively
the same structure; there are also bound singlet $(2e)$--$h$ states above
free LL's (not shown). There are, however, no low-lying bound
singlet $X^-_{s10}$ states.

Due to the selection rule $D({\nu}) \sim \delta_{n,M_z}$,
the $2e$--$h$ states with $M_z=0$ are active in the ExCR transitions
from the $n=0$ LL. As a result, the ExCR transition
$ e^-_0 + {\sl photon} \rightarrow  X^-_{t10}$
to the bound negatively charged triplet MX is prohibited.
ExCR transitions to the bound excited triplet and singlet
$(2e)$--$h$ states are also prohibited: all these states have large $M_z>0$.
Therefore, the ExCR transitions from zero to the first electron LL
can only go to the {\em continuum}. \cite{rmrk2}

All ExCR transitions are only due to LL mixing. In order to
calculate the ExCR dipole transition matrix elements,
I go beyond the high-field limit and admix to (\ref{Psi10})
the triplet $2e$--$h$ states in zero LL's
$\sum_{l=0}^{\infty} \sum_{m=0}^{\infty}
 \gamma_{M_zt\nu}(2m+1,l) \, |000 ; 0 \, 2m+1 \, l \rangle$;
expansion coefficients
$\gamma_{M_zt\nu}(2m+1,l) \sim \delta_{M_z,l-2m-1}$ and
$\gamma_{M_zt\nu} \sim E_0/\hbar\omega_{\rm ce} \sim l_B/a_B$.
The coordinate representation has the form
\begin{eqnarray}
             \nonumber
 & & \Phi_{ml}({\bf r}_1, {\bf r}_2 ;  {\bf r}_h) =
  \langle {\bf r}_1 {\bf r}_2 {\bf r}_h | 000; 0 \, m \, l \rangle  \\
             \label{ml}
 & &   = \frac{1}{\sqrt{l!m!}}
    \left( \frac{z^{\ast}_1 - z^{\ast}_2}{2l_B} \right)^m
    \left( \frac{z_h}{2l_B} \right)^l
    \Phi_{0}({\bf r}_1, {\bf r}_2; {\bf r}_h) \, .
\end{eqnarray}
A similar procedure is used for the singlet
$| \Psi_{M_z s \nu}^{(10)} \rangle$ states.
The dipole transition matrix element is given by
\begin{eqnarray}
       \label{PL2}
 D_t({\nu}) & = & \langle \Psi_{M_z=0 t \nu}| \hat{\cal L}_{\rm PL}
              |\phi^{(e)}_{00} \rangle   \\
         \nonumber
& = & p_{\rm cv} \sum_{m=0}^{\infty}
              \gamma_{M_z=0t\nu}(2m+1,2m+1) {\cal D}_{2m+1} \, , \\
       \label{PL3}
 {\cal D}_{p} & \equiv & \int \! d{\bf r}_1  \! \int \! d{\bf r}_2 \,
   \Phi_{pp}^{\ast}({\bf r}_1, {\bf r}_2; {\bf r}_2)
    \phi^{(e)}_{00}({\bf r}_1)  \, .
\end{eqnarray}
Using Eqs.~(\ref{phi0m}), (\ref{vacuum}), and (\ref{ml})
and performing in (\ref{PL3}) the transformation
$z_2^{\ast}  \rightarrow z_2^{\ast} - z^{\ast}_1$
in the complex $(z_2,z_2^{\ast})$ plane, it can be shown that
the overlap integral ${\cal D}_{p}=(-1)^{p} \, \sqrt{2}$.
The intensity of the ExCR transitions involving the $2e$--$h$
states with eigenenergies $E_{\nu}$ is
\begin{equation}
        \label{PL5}
I^{\rm ExCR}_{s(t)}(\omega) \sim \nu_e
\sum_{\nu} |D_{s(t)}(\nu)|^2 \delta(\hbar\omega-E_{\nu}) \, .
\end{equation}

\begin{figure}[t]
\epsfxsize=3.5in
\epsffile{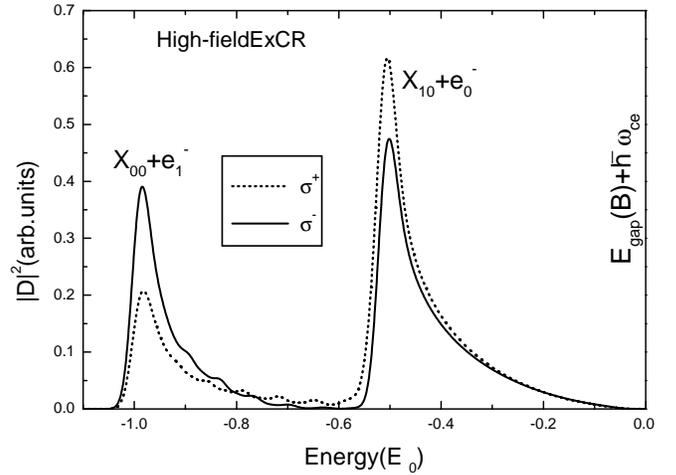}
\vspace*{1ex}
\caption{Energies and dipole matrix elements of the ExCR
transitions from the spin-polarized zero $0$\,$\uparrow$ LL
to the first electron LL in two circular polarizations $\sigma^{\pm}$.
The representative case $E_0=\hbar\omega_{\rm ce}$ is shown.
Spectra have been convoluted with a Gaussian of the $0.015E_0$ width
($\sim 0.3$~meV for the parameters indicated).
}                \label{fig3}
\end{figure}

The calculated dipole  matrix elements of transitions in two circular
polarizations $\sigma^{+}$ and $\sigma^{-}$ that terminate,
respectively, in the continuum of final
singlet $| \Psi_{M_z=0 s \nu}^{(10)} \rangle$
and triplet $| \Psi_{M_z=0 t \nu}^{(10)} \rangle$
states are shown in Fig.~\ref{fig3}.
These ExCR transitions require the extra photon energy
$\sim \hbar\omega_{\rm ce}$ relative to the fundamental
band-gap absorption $E_{\rm gap}(B)$ with final states in the $n=0$ LL.
I neglect the initial and final state exchange self-energy
corrections \cite{FQHE}
$\sim \nu_e E_0 \ll E_0$ that partly compensate each other.
The low-energy ExCR peak in Fig.~\ref{fig3}
corresponds to the transitions
$e_0^- + {\sl photon} \rightarrow X_{00} + e_1^-$
to the lower edge of the continuum formed by the
neutral $1s$ MX in zero LL, $X_{00}$, and the electron
in a scattering state in the first LL.
Such ExCR transitions have been theoretically
identified and discussed in Ref.~\onlinecite{ExCR} in the
low-$B$ limit.

Surprisingly, the present theory predicts in high fields
{\em a new strong feature} --- the second, higher-lying, peak in the ExCR.
If this peak were simply due to the transitions
$e_0^- + {\sl photon} \rightarrow X_{10} + e_0^-$
to a final state consisting of a $2p^+$ neutral MX $X_{10}$,
which is dark in PL, and the electron in a scattering state,
the ExCR transition would be extremely weak and proportional
to the inverse area of the system.
The second ExCR peak may be associated with a formation
of a quasi-bound three-particle state (resonance)
within the $X_{10} + e_0^-$ continuum: The amplitude
of finding all three particles in the same region of real
space is large for a well-defined resonance, so that the
overlap integral [see Eqs.~(\ref{PL2}) and (\ref{PL3})] is large too.
The existence of the $2e$--$h$ resonances is physically
plausible because of the 2D van Hove singularity in the
$X_{10}$ density of states. A well-developed Fano-resonance has
been revealed recently \cite{SSC} in the spectra of intraband internal
$X^-$ transitions, clear evidence supporting
the existence of quasi-bound $2e$--$h$ states.
Note that the predicted double-peak structure in the {\em interband\/}
ExCR transitions physically resembles the predicted \cite{Dz&S_PRL}
and observed \cite{X-int} double-peak
structure in the {\em intraband\/} internal transitions of charged MX's $X^-$.
Although the initial states are different,
both types of spectroscopies probe, in this case, the final
states in the three-particle $2e$--$h$ continuum:
The ExCR final states, as well as the final states in the internal
triplet $X^{-}_{t}$ transitions, have $M_z=0$, while
the final states in the internal singlet
$X^-_{s}$ transitions have $M_z=1$. \cite{Dz&S_PRL,X-int}
This suggests that the ExCR transitions to higher LL's,
may have multiple-peaks and complicated lineshapes
because of more involved structures of the continua.

The intensity of the ExCR lower peak is larger in the $\sigma^-$
polarization (Fig.~\ref{fig3}), which is consistent with
experiment. \cite{ExCR}
The present theory predicts that the intensity of the second ExCR peak
will show the opposite dependence: It should be larger in the
$\sigma^+$ polarization. This polarization dependence
is due to different $e$--$e$ correlations in the final
singlet ($\sigma^+$ polarization) and
triplet ($\sigma^-$ polarization) $2e$--$h$ states:
The oscillator strength is transferred to the higher lying
peak when the final singlet $2e$--$h$ states ---
characterized by a larger $e$--$e$ repulsion ---
are involved.
In agreement with Ref.~\onlinecite{ExCR},
the present theory shows that
(1) the ExCR peaks have intrinsic finite linewidths,
in high fields $\sim 0.15 E_0$ ($\sim 2.6$~meV at $B \simeq 10$~T
corresponding to Fig.~3), and have asymmetric
lineshapes with high-energy tails;
(2) the ExCR transitions are because of LL mixing
and, therefore, ExCR is suppressed in strong fields as
$\nu_e |D|^2 \sim n_e l_B^2 (l_B/a_B)^2 \sim B^{-2}$.

In conclusion, a new magneto-optical phenomenon,
ExCR transitions in low-density 2DEG systems, \cite{ExCR}
has been considered theoretically for high magnetic fields.
The developed formalism allows a consistent treatment of the
final state $e$--$h$ and $e$--$e$ Coulomb interactions.
The features of the high-field ExCR,
in particular, the double-peak structure of the transitions to the first
electron Landau level have been predicted.

I am grateful to B.D.\ McCombe and D.R.\ Yakovlev for useful discussions.
This work was supported in part by a COBASE grant and a grant
of Russian Basic Research Foundation.

\vspace*{-4ex}



\begin{references}
\vspace*{-10ex}

\bibitem[*]{ABD}on leave from General Physics Institute, RAS,
                Moscow 117942, Russia

\bibitem{EP2DS}Proceed. Int. Conf. 
on Electronic Properties of Two-Dimensional Systems 
EP2DS-14  (Prague, Czech Republic,
2001), to be published in Physica E.

\bibitem{Cox}See K.~Kheng {\it et al.},
             Phys. Rev. Lett. {\bf 71}, 1752 (1993);
             A.~J.~Shields {\it et al.},
             Adv. Phys. {\bf 44}, 47 (1995);
             S.~Glasberg {\it et al.},
             Phys. Rev. B {\bf 59}, R10\,425 (1999)
             and references therein.

\bibitem{many-body}E.~I.~Rashba and M.~D.~Sturge,
                   Phys. Rev. B {\bf 63}, 045305 (2000);
                   H.~A.~Nickel {\it et al.,}
                   in Ref.~1 (cond-mat/0106277)
                   and references therein.

\bibitem{Shake}G.~Finkelstein, H.~Shtrikman, and I.~Bar-Joseph,
               Phys. Rev. B {\bf 53}, 12\,593 (1996);
               {\bf 56}, 10\,326 (1997)
               and references therein.

\bibitem{ExCR}D.~R.~Yakovlev {\it et al.,}
              Phys. Rev. Lett. {\bf 79}, 3974 (1997).

\bibitem{Simon}J.~E.~Avron, I.~W.~Herbst, and B.~Simon,
               Ann. Phys. (N.Y.) {\bf 114}, 431 (1978).

\bibitem{Dz&S_PRL}A.~B.~Dzyubenko and A.~Yu.~Sivachenko,
                  Phys. Rev. Lett. {\bf 84}, 4429 (2000).

\bibitem{SSC} A.~B.~Dzyubenko, Solid State Commun. {\bf 113},
              683 (2000); 
              Phys. Rev. B in press; cond-mat/0107611.

\bibitem{rmrk}This is quantitatively correct at low magnetic fields,
when the bound complex rotates in $B$ as a whole.

\bibitem{FQHE}See, e.g., Z.~F.~Ezawa, {\it Quantum Hall Effects\/}
               (World Scientific, Singapore, 2000).

\bibitem{L&L80}I.~V.~Lerner and Yu.~E.~Lozovik,
               Sov. Phys. JETP {\bf 51}, 588 (1980).

\bibitem{rmrk2}
This conclusion is reached for a strictly-2D system in the limit of
high magnetic fields. It is applicable to quasi-2D systems at finite
fields --- as long as no bound $X^-$ states with $M_z=0$ appear
in the first electron LL.


\bibitem{X-int}H.~A.~Nickel {\it et al.,}
               Phys. Stat. Sol. B {\bf 210}, 341 (1998);
                 A.~B.~Dzyubenko {\it et al.,}
                 Physica E {\bf 6}, 156 (2000).

\end{references}
\end{document}